# Millikelvin cooling of an optically trapped microsphere in vacuum

Tongcang Li, Simon Kheifets, and Mark G. Raizen

Center for Nonlinear Dynamics and Department of Physics, The University of Texas at Austin, Austin, TX 78712, USA

The apparent conflict between general relativity and quantum mechanics remains one of the unresolved mysteries of the physical world[1,2]. According to recent theories[2-4], this conflict results in gravity-induced quantum state reduction of "Schrödinger cats", quantum superpositions of macroscopic observables. In recent years, great progress has been made in cooling micromechanical resonators towards their quantum mechanical ground state[5-12]. This work is an important step towards the creation of Schrödinger cats in the laboratory, and the study of their destruction by decoherence. A direct test of the gravity-induced state reduction scenario may therefore be within reach. However, a recent analysis shows that for all systems reported to date, quantum superpositions are destroyed by environmental decoherence long before gravitational state reduction takes effect[13]. Here we report optical trapping of glass microspheres in vacuum with high oscillation frequencies, and cooling of the center-of-mass motion from room temperature to a minimum temperature of 1.5 mK. This new system eliminates the physical contact inherent to clamped cantilevers, and can allow ground-state cooling from room temperature[14-20]. After cooling, the optical trap can be switched off, allowing a microsphere to undergo free-fall in vacuum[20]. During free-fall, light scattering and other sources of environmental decoherence are absent, so this system is ideal for studying gravitational state reduction[21]. A cooled optically trapped object in vacuum can also be used to search for non-Newtonian gravity forces at small scales[22], measure the impact of a single air molecule[19], and even produce Schrödinger cats of living organisms[14].

Previous experiments demonstrated optical levitation of a 20-μm diameter sphere in vacuum with a trapping frequency of about 20 Hz, as well as feedback control of a trapped sphere which was used to increase the trapping frequency to several hundred hertz and stabilize its position to within a fraction of one micrometer[23, 24]. However, the resolution of its detection system[24] was not sufficient to enable feedback cooling. According to the equipartition theorem, the root mean square (rms) amplitude of a trapped microsphere at thermal equilibrium is $x_{rms} = \sqrt{k_B T_0/(m\omega^2)}$, where $k_B$ is the Boltzmann constant, $T_0$ the air temperature, $m$ the mass of the microsphere, and $\omega$ the angular trapping frequency. For a 20-μm diameter sphere trapped at one hundred hertz, the rms amplitude is about 0.04 μm at 300 K, and will be much smaller at lower temperature. It is also important that the trapping frequency be much higher than the frequencies of seismic vibration in order to achieve significant cooling.

We use a dual-beam optical tweezer to trap a 3.0-μm diameter $SiO_2$ sphere in vacuum with much higher oscillation frequencies (about 10 kHz) to minimize the effects of instrumental vibration. We also demonstrate a detection system to monitor the motion of a trapped microsphere with a sensitivity of about 39 fm/Hz$^{1/2}$ over a wide frequency range. Using active feedback, we simultaneously cool the three center-of-mass (COM) vibration modes of a microsphere from room temperature to milikelvin temperatures with a minimum mode temperature of 1.5 mK, which corresponds to the reduction of the rms amplitude of the microsphere from 6.7 nm to 15 pm for that mode.

A simplified scheme of our optical trap and feedback setup is shown in Fig. 1. The optical trap is similar to a trap used to measure the instantaneous velocity of a Brownian particle in air[25]. It is created inside a vacuum chamber by two counter-propagating laser beams focused to the same point by two identical aspheric lenses with a focal length of 3.1 mm and numerical aperture of 0.68. The wavelength of both trapping beams is 1064 nm. They are orthogonally polarized, and are shifted in frequency to avoid interference. The beams are slightly elliptical and approximately form a harmonic trap with three fundamental vibration modes along the horizontal, vertical and axial directions, denoted X, Y, and Z in Fig. 1a. The motion of a trapped bead causes deflection of both trapping beams. We monitor the position of the bead by measuring the deflection of one of the trapping beams with ultrahigh spatial and temporal resolution in all three dimensions (Supplementary Figure 1).

Using the position signal, we can calculate the instantaneous velocity of the bead, and implement feedback cooling by applying a force with a direction opposing the velocity (Fig. 1b). The feedback is generated by scattering forces from three orthogonal 532 nm laser beams along the axes as shown in Fig. 1a. The average intensity of the cooling beams is about 1% of the trapping beams. The optical power of each cooling beam is controlled by an acousto-optic modulator (AOM). Each beam is modulated with a time-varying signal proportional to the instantaneous velocity of the bead, added to an offset. The proportional component generates the required cooling force, while the offset slightly shifts the trap center.

An optically trapped microsphere in non-perfect vacuum will exhibit Brownian motion due to collisions between the microsphere and residual air molecules. When the microsphere is at thermal equilibrium, the power spectrum of COM motion along each of the three fundamental mode axes is[9,26]:

$$S_j(\Omega) = \frac{2k_B T_0}{m} \frac{\Gamma_0}{(\omega_j^2 - \Omega^2)^2 + \Omega^2 \Gamma_0^2} \quad (1)$$

where $\Omega/2\pi$ is frequency, $\Gamma_0$ is the viscous damping factor due to the air, and $\omega_j$ (j=1, 2, 3) are the resonant frequencies of the optical trap along the x, y, and z axes.

The viscous damping factor due to air can be calculated by kinetic theory. Assuming the reflection of air molecules from the surface of a microsphere is diffusive, and the molecules thermalize with the surface during collisions, we obtain[27]:

$$\Gamma_0 = \frac{6\pi\eta r}{m} \frac{0.619}{0.619 + Kn}(1 + c_K) \quad (2)$$

where η is the viscosity coefficient of the air, $r$ the radius of the microsphere, and $Kn = s/r$ the Knudsen number. Here $s$ is the mean free path of the air molecules. $c_K = (0.31Kn)/(0.785+1.152Kn+Kn^2)$ is a small positive function of $Kn$. At low pressures where $Kn \gg 1$, the viscous damping factor is proportional to the pressure.

The behavior of the system with three dimensional (3D) feedback cooling is straightforward to understand if we assume that there is no coupling between feedback forces and velocities in different directions. In this case, the feedback force in each direction adds an effective cold damping factor $\Gamma^{fb}_j$, and the total damping becomes $\Gamma^{tot}_j = \Gamma_0 + \Gamma^{fb}_j$. The power spectrum of the motion of a trapped microsphere with feedback cooling can be described by Eq. 1 with $T_0$ and $\Gamma_0$ replaced by $T^{fb}_j$ and $\Gamma^{tot}_j$, where $T^{fb}_j = T_0\Gamma_0/\Gamma^{tot}_j$ is the temperature of the motion with feedback cooling[9]. The motion can be cooled significantly by applying feedback damping $\Gamma^{fb}_j \gg \Gamma_0$. The lowest temperature will be limited by the noise in the detection system and feedback circuits, as well as coupling between different directions.

Figure 2 shows the linewidths, $\Gamma_0/2\pi$, of the oscillation of a trapped 3-μm microsphere at different pressures without feedback cooling. The powers of the two trapping beams are 120 mW and 100 mW, respectively. The linewidths are obtained by fitting the measured power spectra with Eq. 1. The measured linewidths agree very well with the prediction of kinetic theory (Eq. 2) from $10^5$ Pa down to 1 Pa. At pressures below 1 Pa, the measured linewidths are larger than the theoretical prediction. This linewidth broadening is due to power fluctuations of the trapping laser. The inset of Fig. 2 shows a power spectrum at 0.13 Pa. The trapping frequency $\omega_1/2\pi$ is 9756.4±0.3 Hz, and the linewidth is 0.46 ±0.06 Hz, giving a quality factor ($Q_j = \omega_j/\Gamma_0$) of $2.1 \times 10^4$. This implies the power fluctuation of the trapping laser is smaller than 0.01% during the measurement. An optically trapped microsphere provides a method to directly convert laser power to a frequency signal, which can be measured precisely. Stabilization of laser power to a trapped bead can find applications in laser physics, and can enable a more precise measurement of the Q for a second trapped bead in vacuum.

Figure 3 shows experimental results of feedback cooling. Before feedback is turned on, the resonant frequencies $\omega_j/2\pi$ are 8066±5 Hz, 9095±4 Hz, and 2072±6 Hz for the fundamental modes at 637 Pa along the x, y, and z axes, respectively. At this pressure, the peaks in the power spectrum due to the three fundamental modes are distinguishable, and heating effects due to the laser are negligible. We can therefore use the measured power spectra at 637 Pa to calibrate the position detectors for the fundamental modes at room temperature. After we turn on feedback cooling, the temperature of the y mode changes from 297 K to 24 K at 637 Pa. Then we reduce the air pressure while keeping the feedback gain almost constant, thus the heating rate due to collisions from air molecules decreases, while the cooling rate remains constant. As a result, the temperature of the motion drops. At 5.2 mPa, the mode temperatures are 150±8 mK, 1.5±0.2 mK, and 68±5 mK for the x, y and z modes. The mean thermal occupancy $<n> = k_B T^{fb}_j/(\hbar\omega_j)$ of the y mode is reduced from about $6.8\times10^8$ at 297K to about 3400 at 1.5 mK. Here $\hbar$ is the reduced Planck constant. Fig. 3d shows the temperature of the three fundamental modes as a function of pressure. At low pressure and when the feedback gain is constant, the mode temperature should be proportional to the pressure, which is shown as a straight line with slope 1 in the figure. The temperature of the y mode agrees with this prediction very well at pressures above 1 Pa.

At our lowest temperatures, the power spectra are still much larger than the noise level, and the minimum temperature is achieved at pressures above the minimum pressure we can obtain, thus the electronic noise (in detection and feedback circuits) and the pressure are currently not the limiting factor of the experiment. The dominant limiting factor is most likely residual coupling between intensity and direction of the cooling beams. When we change the intensity of a cooling beam by an AOM, the direction and profile of the beam is also changed slightly. This causes heating of the motion of a microsphere perpendicular to the beam while cooling it parallel. This problem should be solved by replacing the AOM's with electro-optic modulators (EOM's). With feedback cooling, we have trapped a microsphere for more than one hour at pressure below 1 mPa. This lifetime should be long enough to perform cavity cooling[14,15,19]. We have been able to charge the microsphere by high voltage electrical breakdown of nearby air. After charging, we use electrostatic forces to implement feedback cooling, and have been able to cool the motion of a trapped microsphere to about 10 K. This provides a method for combining an optical trap and an ion trap at one place, which will help to trap and study particles at ultrahigh vacuum.

Our result is an important step toward quantum ground-state cooling of a trapped macroscopic object in vacuum by either cavity cooling[14,15,19] or feedback cooling with an improved detection and feedback scheme[28,29]. For cavity cooling of a trapped object in vacuum, it is also important to use feedback cooling to pre-cool and stabilize the object, in order to have enough time to tune the cavity cooling laser to correct frequency for efficient cooling. After cooling and creation of a superposition state in momentum, the optical trap can be switched off to let a microsphere undergo free-fall in vacuum[20]. The wavefuction will expand during free-fall and become a superposition state in space[21]. The finite lifetime of a superposition due to gravity-induced state reduction is predicted to be of the order of $\hbar r/(Gm^2)$ when the superposition is composed of states separated by a distance larger than the size of the microsphere[2,13], where $G$ is Newton's gravitational constant. The predicted lifetime is about 3 ms for a 3-μm diameter microsphere, which is much shorter than the environmental decoherence time in good vacuum and thus measurable.

---

1. Hawking, S. W. & Isreal, W. (eds.) *General Relativity; an Einstein Centenary Survey* (Cambridge University Press, Camberidge, 1979).

2. Penrose, R. On gravity's role in quantum state reduction. *Gen. Rel. Grav.* **28**, 581-600 (1996).

3. Diósi, L. Models for universal reduction of macroscopic quantum fluctuations. *Phys. Rev. A* **40**, 1165-1174 (1989).

4. Ghirardi, G. C. Rimini, A. & Weber, T. Unified dynamics for microscopic and macroscopic systems. *Phys. Rev. D* **34**, 470-491 (1986).

5. Metzger, C. H. & Karrai, K. Cavity cooling of a microlever. *Nature* **432**, 1002-1005 (2004).

6. Naik, A. *et al*. Cooling a nanomechanical resonator with quantum back-action. *Nature* **443**, 193-196 (2006).

7. Gigan, S. *et al*. Self-cooling of a micromirror by radiation pressure. *Nature* **444**, 67-70 (2006).

8. Arcizet, O. Cohadon, P.-F. Briant, T. Pinard, M. & Heidmann, A. Radiation-pressure cooling and optomechanical instability of a micromirror. *Nature* **444**, 71-74 (2006).

9. Kleckner, D. & Bouwmeester, D. Sub-kelvin optical cooling of a micromechanical resonator. *Nature* **444**, 75-78 (2006).

10. Schliesser, A. Del'Haye, P. Nooshi, N. Vahala, K. J. & Kippenberg, T. J. Radiation pressure cooling of a micromechanical oscillator using dynamical backaction. *Phys.Rev. Lett.* **97**, 243905 (2006).

11. Thompson, J. D. *et al*. Strong dispersive coupling of a high-finesse cavity to a micromechanical membrane. *Nature* **452**, 72-75 (2008).

12. O'Connell, A. D. *et al*. Quantum ground state and single-phonon control of a mechanical resonator. *Nature* **464**, 697-703 (2010).

13. van Wezel, J. Oosterkamp, T. & Zaanen J. Towards an experimental test of gravity-induced quantum state reduction. *Phil. Mag.* **88**, 1005-1026 (2008).

14. Romero-Isart, O. Juan, M. L. Quidant, R. & Cirac, J. I. Toward quantum superposition of living organisms. *N. J. Phys*. **12**, 033015 (2010).

15. Chang, D. E. *et al*. Cavity opto-mechanics using an optically levitated nanosphere. *Proc. Natl. Acad. Sci. USA* **107**, 1005-1010 (2010).

16. Singh, S. Phelps, G. A. Goldbaum, D. S. Wright, E. M. & Meystre, P. All-optical optomechanics: an optical spring mirror. *Phys. Rev. Lett.* **105**, 213602 (2010).

17. Barker, P. F. Doppler cooling a microsphere. *Phys. Rev. Lett.* **105**, 073002 (2010).

18. Schulze, R. J. Genes, C. & Ritsch, H. Optomechanical approach to cooling of small polarizable particles in a strongly pumped ring cavity. *Phys. Rev. A* **81**, 063820 (2010).

19. Yin, Z. Q. Li, T. & Feng, M. Three dimensional cooling and detecting of a nanosphere with a single cavity. http://arxiv.org/abs/1007.0827 (2010).

20. Romero-Isart, O. *et al*. Optically levitating dielectrics in the quantum regime: theory and protocols. *Phys. Rev. A* **83**, 013803 (2011).

21. Romero-Isart, O. *et al*. in preparation.

22. Geraci, A. A. Papp, S. B. & Kitching, J. Short-range force detection using optically cooled levitated microspheres. *Phys. Rev. Lett.* **105**, 101101 (2010).

23. Ashkin, A. & Dziedzic, J. M. Optical levitation in high vacuum. *Appl. Phys. Lett*. **28**, 333-335 (1976).

24. Ashkin, A. & Dziedzic, J. M. Feedback stabilization of optically levitated particles. *Appl. Phys. Lett.* **30**, 202-204 (1977).

25. Li, T. Kheifets, S. Medellin, D. & Raizen, M.G. Measurement of the instantaneous velocity of a Brownian particle. *Science* **328**, 1673-1675 (2010).

26. Berg-Sørensen, K. & Flyvbjerg, H. Power spectrum analysis for optical tweezers. *Rev. Sci. Instrum.* **75**, 594-612 (2004)

27. Beresnev, S. A. Chernyak, V. G. & Fomyagin, G. A. Motion of a spherical particle in a rarefied gas. Part 2. Drag and thermal polarization. *J. Fluid Mech.* **219**, 405-421 (1990).

28. Mancini, S. Vitali, D. & Tombesi, P. Optomechanical cooling of a macroscopic oscillator by homodyne feedback. *Phys. Rev. Lett.* **80**, 688-691 (1998).

29. Genes, C. Vitali, D. Tombesi, P. Gigan, S. & Aspelmeyer, M. Ground-state cooling of a micromechanical oscillator: comparing cold damping and cavity-assisted cooling schemes. *Phys. Rev. A* **77**, 033804 (2008).

**Acknowledgements** M.G.R acknowledges support from the Sid W. Richardson Foundation and the R. A. Welch Foundation, grant number F-1258. T.L. is supported by the Lawrence C. Biedenharn Endowment for Excellence. We thank O. Romero-Isart, D. Medellin, and Z. Q. Yin for helpful discussions.

**Author Information** Correspondence and requests for materials should be addressed to M. G. R. (raizen@physics.utexas.edu).

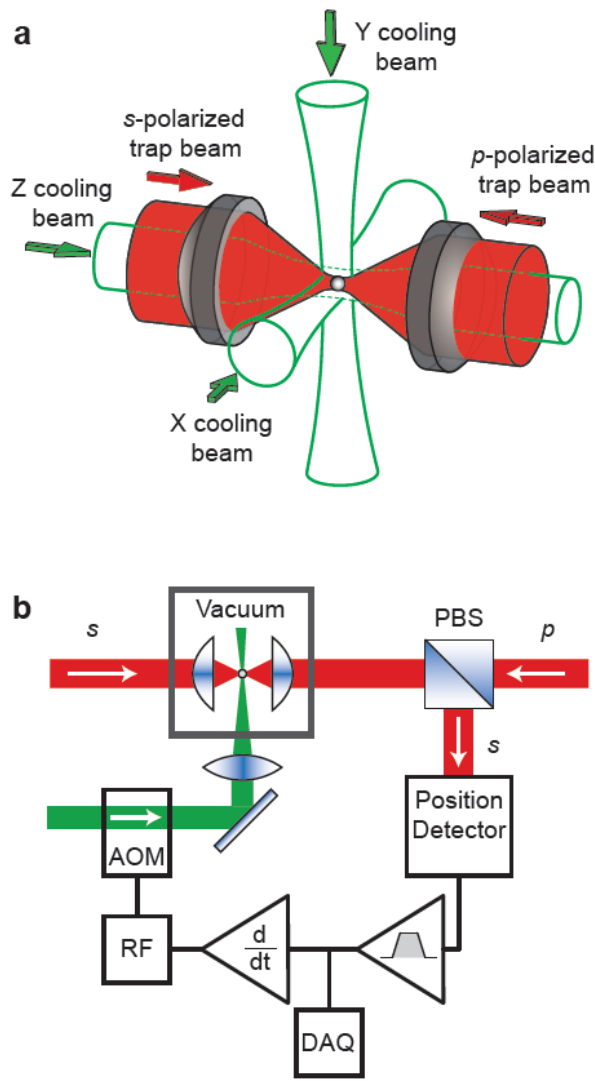

**Figure 1 | Schematic of feedback cooling of an optically trapped microsphere. a**, Simplified schematic showing a glass microsphere trapped at the focus of a counter-propagating dual-beam optical tweezer, with three laser beams along the axes for cooling. The wavelengths of the trapping beams and the cooling beams are 1064 nm and 532 nm, respectively. **b**, Diagram of the feedback mechanism for the X axis: The position of a trapped microsphere is monitored by a home-built detecting system. The position signal is sent through a bandpass filter (typically 100 Hz to 300 kHz) and a derivative circuit (d/dt) to provide a signal proportional to velocity. This velocity signal is used to control the output power of a radio frequency (RF) AOM driver which modulates the power of the X cooling beam. The data is digitized and stored on a computer by a data acquisition card (DAQ).

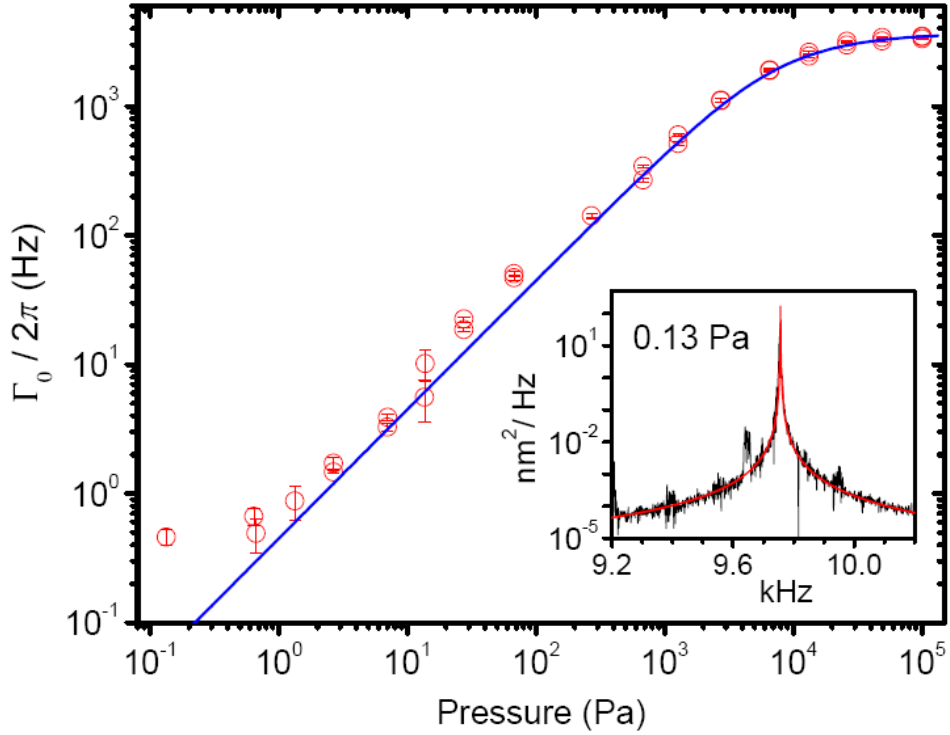

**Figure 2 | Measured linewidths of the oscillation of an optically trapped 3-μm diameter microsphere at different pressures.** The blue curve is the prediction of a kinetic theory (Eq. 2). The inset is the measured power spectrum at 0.13 Pa. By fitting the spectrum with Eq. 1 (red curve), we obtain $\omega_1 = 2\pi$ (9756.4 ±0.3) Hz, and $\Gamma_0 = 2\pi$ (0.46 ±0.06) Hz for this example. The same method is used to obtain linewidths for other pressures.

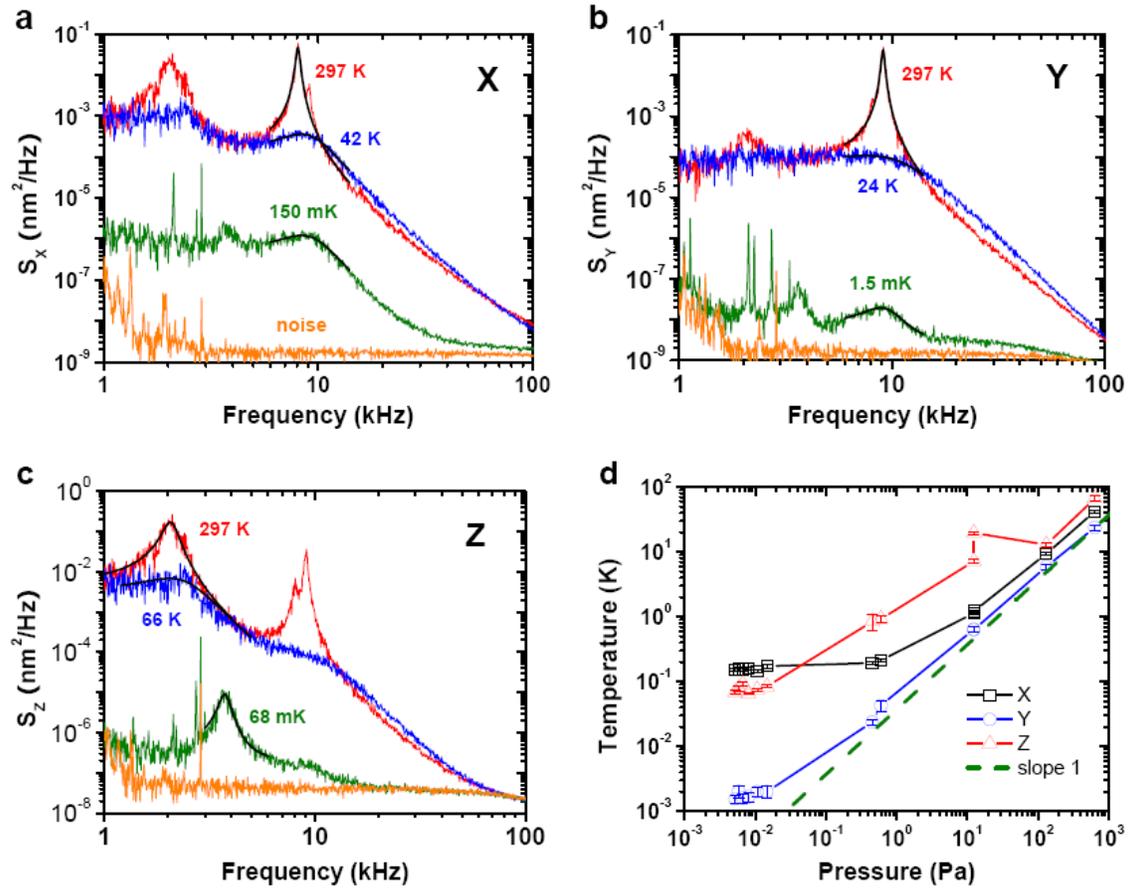

**Firgure 3 | Experimental results of feedback cooling. a, b, and c** show the power spectra of a trapped 3 μm diameter microsphere along the X, Y, and Z axes as it is cooled. Red curves are intrinsic spectra at 637 Pa without feedback cooling, blue curves are spectra at 637 Pa with feedback cooling, green curves are spectra at 5.2 mPa with feedback cooling, and orange curves are the noise signals when there is no particle in the optical trap. The black curves are fits of a thermal model (see text for details). We obtain mode temperatures from these fits. **d**, Temperatures of the three fundamental oscillation modes along X (black squares), Y (blue circles), and Z (red triangles) axes as a function of the air pressure. The dashed line is a straight line with slope 1 for comparison.

# Supplementary material for "Millikelvin cooling of an optically trapped microsphere in vacuum"

Tongcang Li, Simon Kheifets, and Mark G. Raizen

Center for Nonlinear Dynamics and Department of Physics, The University of Texas at Austin, Austin, TX 78712, USA

**Three dimensional detection system**

The Supplementary Figure 1 shows the detection system used to monitor the real-time position of a trapped microsphere with ultra-high precision in 3D. The X, Y, and Z detectors are fast balanced photo-detectors with bandwidth of 75MHz. They have two matched photodiodes to cancel the common mode noise in the laser beams, allowing ultra-high precision measurements of the position of a trapped microsphere. When a trapped microsphere moves in the horizontal (vertical) direction, it deflects the trapping beam in the horizontal (vertical) direction. This changes the relative power between the two beams after the MX (MY) mirror, which is measured by the X (Y) detector. The motion of a trapped bead along the trap axis changes the divergence angle of the output beam, which changes the waists of incident beams on the Z detector. One photodiode of the Z detector is smaller than the waist of the incident beam. It measures only part power of the incident beam. Thus its output voltage depends on the waist of the incident beam, which is a function of the position of the microsphere in Z axis. The other photodiode of the Z detector is much larger than the waist of the incident beam. It measures the total power of the incident beam. Thus its output voltage does not depend on the waist of the incident beam, and can serve as a reference signal.

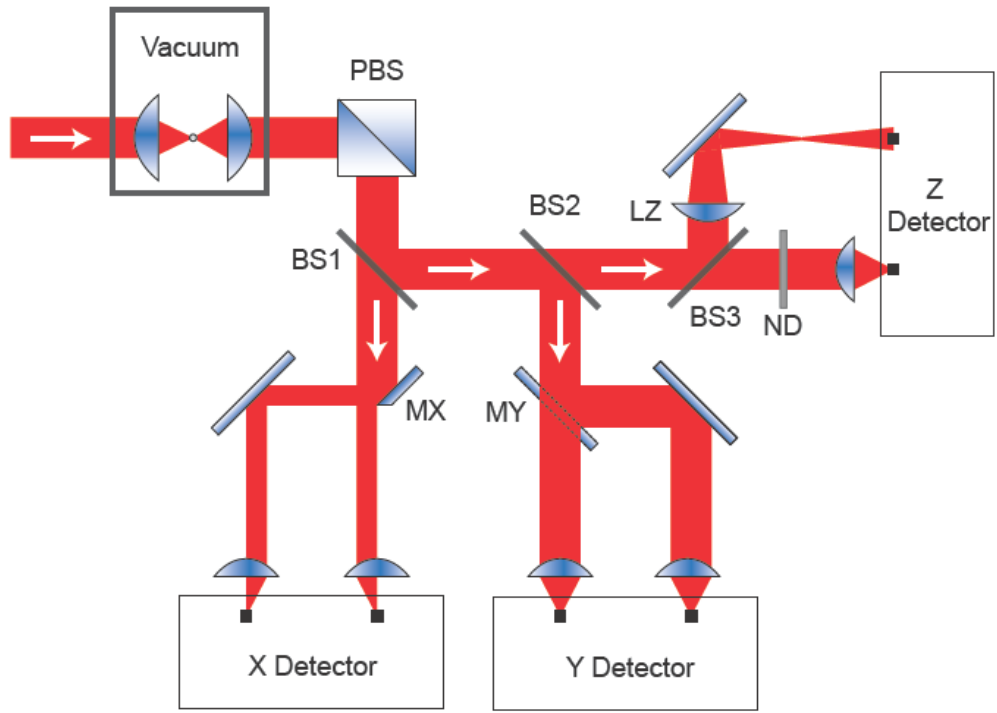

**Supplementary Figure 1**: Simplified schematic showing the detection system that monitors the real-time position of a trapped microsphere with ultra-high precision in all three dimensions. One of the trapping beams (the other trapping beam is not shown) passes through a trapped microsphere inside a vacuum chamber and is reflected by a polarizing beam splitter cube (PBS). It is then split to three beams by two beam splitters (BS1 and BS2) for 3D detection. MX is a mirror with a sharp edge that splits the beam into two parts horizontally. MY is a mirror with a sharp edge that splits the beam into two parts vertically. BS3 is a beam splitter, ND is a neutral density filter, and LZ is a lens. The X, Y, and Z detectors are balanced detectors that have two matched photodiodes to cancel the common mode noise in the laser beams.

**Calibration of the detection system**

For small displacements of a trapped microsphere near the trap center, the voltage output ($U_i$) of each detector is proportional to the displacement ($x^D_i$) of the microsphere along the detection direction, i.e. $U_i = \beta_i x^D_i$, where $i = 1, 2, 3$ denotes X, Y, and Z detectors, and $\beta_i$ the calibration factor of the detector. We align the detection system carefully to make the detection direction of each detector be parallel to one of the trap's fundamental mode axes. In reality, however, there is always slight difference between the detection directions and the fundamental mode axes. Thus the voltage output from each detector is a combination of signals from each mode, i.e. $U_i = \beta_i(\alpha_{i1} x^M_1 + \alpha_{i2} x^M_2 + \alpha_{i3} x^M_3)$, where $x^M_j$ ($j = 1, 2, 3$) is the displacement of the microsphere along the fundamental mode axis $\hat{x}^M_j$, and $\alpha_{ij}$ is the projection coefficient of $\hat{x}^M_j$ to the detection direction $\hat{x}^D_i$. Usually only one term dominates as the detection directions are almost parallel to the mode axes.

The expected value of the power spectrum of the voltage output from each detector is[25]:

$$S^U_i(\Omega) \equiv \left\langle |\tilde{U}_i|^2 / T_{msr} \right\rangle = \beta_i^2 <| \alpha_{i1}\tilde{x}^M_1 + \alpha_{i2}\tilde{x}^M_2 + \alpha_{i3}\tilde{x}^M_3 |^2 / T_{msr} >$$

where $T_{msr}$ is the measurement time, $\tilde{U}_i$ and $\tilde{x}^M_j$ are Fourier transforms of $U_i$ and $x^M_j$, respectively. The expansion of $S^U_i(\Omega)$ has 9 terms, but the expected values of the cross correlation terms $<\tilde{x}^M_i \tilde{x}^M_j>$ ($i \neq j$) are 0 because the movement of a microsphere in three dimensions of a harmonic trap are uncorrelated. Thus

$$S^U_i(\Omega) = \beta_i^2 [\alpha_{i1}^2 S_1(\Omega) + \alpha_{i2}^2 S_2(\Omega) + \alpha_{i3}^2 S_3(\Omega)]$$

where $S_j(\Omega)$ is the power spectrum of COM motion along each fundamental mode axis. We have

$$S_j(\Omega) = \frac{2k_B T_0}{m} \frac{\Gamma_0}{(\omega_j^2 - \Omega^2)^2 + \Omega^2 \Gamma_0^2}$$

when there is no feedback cooling (the same as Eq. (1) in the main text); and

$$S_j(\Omega) = \frac{2k_B T_j^{fb}}{m} \frac{\Gamma_j^{tot}}{(\omega_j^2 - \Omega^2)^2 + \Omega^2 (\Gamma_j^{tot})^2}$$

with feedback cooling.

The detection system can be calibrated by fitting the measured power spectra at room temperature with the expected power spectra $S_i^U(\Omega)$ to obtain calibration factor $\beta_i^2 \alpha_{ij}^2$ for each mode that is distinguishable in the power spectra. We can also obtain $\beta_i^2$ directly by the energy equipartition theorem, which says $<mv^2_i/2> = k_B T_0/2$, where $v_i$ is the instantaneous velocity of the microsphere projected onto any axis. Since $U_i = \beta_i x^D_i$, we have $\beta_i^2 = \frac{m}{k_B T_0} <\left(\frac{dU_i}{dt}\right)^2>$. With $\beta_i^2 \alpha_{ij}^2$ and $\beta_i^2$, we can easily obtain $\alpha_{ij}^2$ to check the alignment of our detection system. In the experiment, each detector is used to monitor only one mode, so only three calibration factors $\beta_1^2 \alpha_{11}^2$, $\beta_2^2 \alpha_{22}^2$, and $\beta_3^2 \alpha_{33}^2$) are required for measuring the mode temperatures with feedback cooling.

The mass of the microsphere is required to obtain the calibration factors. The pure silica (SiO$_2$) microspheres used in this experiment are from Bangs Laboratories, Inc. Their mean diameter is 3.0 μm, corresponding to a mean mass of $2.8 \times 10^{-14}$ kg for each microsphere. The standard deviation of the size given by the supplier is 14 %. The exact diameter of the microsphere is not important for feedback cooling. The temperatures of the feedback-cooled motion are obtained by comparing the power spectra of the same microsphere with and without feedback cooling, a measurement which is independent of the exact size of the microsphere. The viscous damping factor ($\Gamma_0$) of a microsphere in air, however, depends on the size of the microsphere. Using the measured damping factor shown in Fig. 2 of the main text, we obtain the diameter of a microsphere by kinetic theory (Eq. 2) to be 2.7 μm, which is within the uncertainty range given by the supplier. All data shown in Fig. 2 is from one microsphere, and all data shown in Fig. 3 is from another microsphere.